\documentclass[11pt,dvips]{article}
\usepackage{epsfig,times}
\usepackage{floatflt}
\usepackage{picinpar}
\usepackage{moriond}
\usepackage{a4}

\graphicspath{{./plots/},{.}}
\begin{document}
\title{Observations with the HEGRA stereoscopic system}
\author{D. Horns for the HEGRA collaboration}
\address{Max-Planck-Institut f\"ur Kernphysik, Postfach 10\,39\,80, D-69029
Heidelberg, Germany}
\maketitle
\abstracts{
 The HEGRA system of imaging air Cherenkov telescopes has successfully pioneered
the stereoscopic observation technique of extensive air showers. The observational method
is briefly described and important results of recent observations of 
sources of photons with TeV($=10^{12}$ eV)-energies are summarized:
 The first detection of a TeV $\gamma$-ray signal from the shell-type supernova remnant 
Cassiopeia A and preliminary results obtained from the observation of strong variability of the extragalactic source Mkn 421 during
observations carried out from February to May 2000.
}

\section{Introduction}
 The sensitivity for detection of TeV-$\gamma$-ray sources 
by observation of extensive air showers in the atmosphere has been largely improved 
with respect to previously used methods by
introducing the air Cherenkov \textit{imaging} technique \cite{Whipple}.
 With this technique, it is possible to
reduce the number of background events induced by charged cosmic rays 
by applying cuts on the orientation and shape of images 
in the focal plane of imaging air Cherenkov telescopes (IACTs).
 By observing air showers simultaneously with more than one telescope the
shower geometry can be reconstructed with an accuracy that is unmatched in the
field of air shower detectors. The HEGRA collaboration has 
pioneered successfully the stereoscopic technique with the system of five IACTs operational since 1997 \cite{Daum97}
on the Canary island La Palma (2200 m asl, 17.89$^\circ$ W, 27.9$^\circ$N).
\section{Stereoscopic observation method}
 \subsection{Detector setup and data taking}
 The HEGRA IACTs are placed on the corners of a square with a side length of
roughly 100 m with one IACT in the center. The IACTs each  
have a tessellated mirror with a collection area of $8.5$
m$^2$ and a focal length of 5 m. In the prime focus a matrix of 271
photomultiplier tubes is installed, detecting the faint flashes of air
Cherenkov light produced by charged particles of extensive air showers.  The
telescopes observe co-aligned a patch in the sky with a diameter of
$4.3^\circ$ with an absolute pointing accuracy of $\approx 0.007^\circ$.

 The data are taken in moonless clear nights with  a mean event rate of 15-18
Hertz for observations carried out close to the zenith. The peak detection
rate for a photon source in the zenith with an energy spectrum similar to
the one from the Galactic standard candle (Crab Nebula) is at an energy of
500 GeV.  Images of all telescopes are read out after two telescopes have
been triggered by Cherenkov light from an air shower within a time window of
70 ns \cite{Bulian}.  \subsection{Reconstruction of shower geometry} The
images from the single cameras are cleaned by a \textit{tailcut}, removing
pixel amplitudes that are due to sky noise. Images with an integral
amplitude of more than 40 photo electrons are parameterized by the first and
second moments of the pixel amplitude distribution. The average of the pairwise
intersections of the major axes of the image ellipsoids 
is calculated to determine the direction of incidence of the air shower.
Several schemes have been developed to improve 
the angular \mbox{resolution
\cite{hofmann}}. The different methods achieve accuracies ranging from
$\theta_{res}=0.05^\circ \,\mbox{to}\, 0.1^\circ$ for photon-induced
showers. The intersection point of the shower axis with the plane that is
perpendicular to the optical axis of the telescopes (\textit{shower core})
is calculated in a similar way by determining the intersection point of the
major axes of the image ellipsoids in a different reference frame. Depending
upon the method chosen \cite{hofmann} resolutions of \mbox{2-10 m} for the
position of the shower core  are achievable. 
 \subsection{Separation of
$\gamma$-and hadron-induced air showers} The detection of $\gamma$-ray
sources with the air shower technique is made difficult by background events
induced by charged cosmic rays. Since the charged cosmic rays have an
isotropic distribution of  arrival directions, the background can be
suppressed for a point source by a factor that is roughly proportional to
$\theta_{res}^2/\Omega_{fov}$, where $\theta_{res}$ is the angular
resolution and $\Omega_{fov}$ is the solid angle covered by the field of
view. The imaging technique allows to suppress the background even further
by selecting $\gamma$-like events by cutting on the shape of the image.
Again, the stereoscopic approach allows for an improvement of the capability
to separate $\gamma$-induced showers from charged cosmic ray events: The
\textit{width} of the image ellipsoids is scaled to the expectation value
for the image width derived for photon-induced showers using the apparent
image brightness and the core position.  The average of the scaled width
over all telescopes is then used as a cut parameter. In this way
 hadron-induced air shower events can be rejected to the relative level of a few per
cent while retaining 60\,\% of the photon-induced showers \cite{konopelko}.
%\texttt{energy resolution, flux sensitivity, detector description, lateral
%distribution, shower maximum etc.}
\section{First detection of a TeV signal from Cassiopeia A}
\begin{figwindow}[6,r,{ \includegraphics[width=5cm]{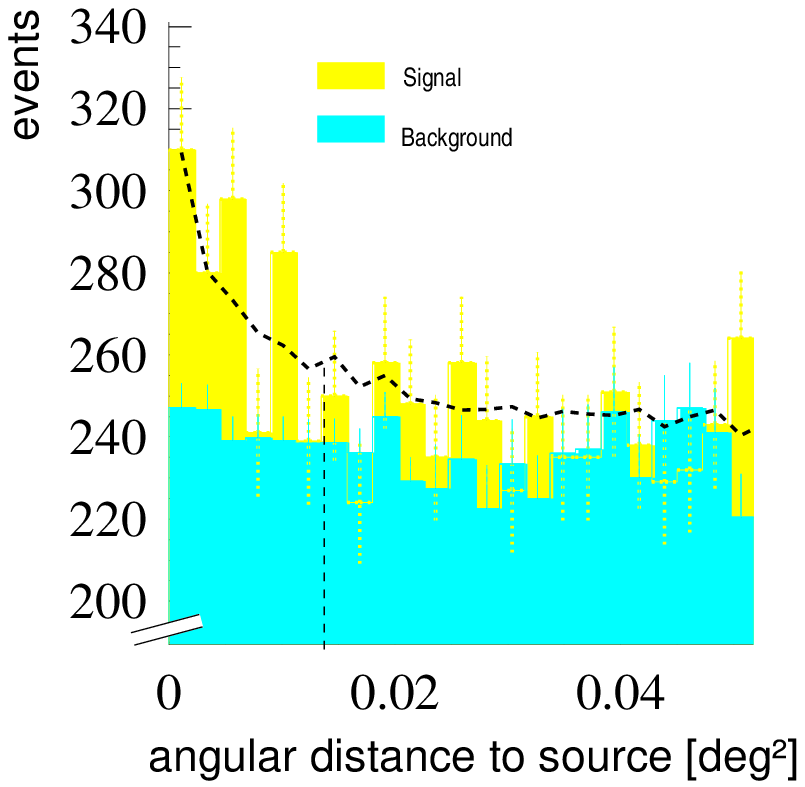}}, { As a
function of squared angular distance, the signal from the ON-region is seen
above the flat background estimated from the OFF-region. The excess follows
the expected shape indicated by the dashed curve.\label{fig:signal}}] The
young shell-type supernova remnant \textit{Cassiopeia A} (Cas A) is an
exceptionally bright object in the X-ray and radio wavelength bands.  A
large fraction of this emission of the nearby remnant (3.3 kpc) is
attributed to synchrotron radiation from a population of relativistic
electrons \cite{atoyan_casa}. The same electron population, which is
presumably accelerated in the vicinity of the shock formed at the boundary
between the expanding shell and the surrounding interstellar medium, could
produce TeV photons via Compton scattering processes and Bremsstrahlung in
the ambient interstellar medium (leptonic origin). Another  
production mechanism for TeV photons could be due to accelerated nucleons
which would produce neutral pions in interaction processes with the
interstellar medium.  
 The neutral pions decay into observable TeV photons
(nucleonic origin).  If the nucleonic
origin is verified, this would identify for the first time  an important
source of Galactic cosmic rays.
Motivated by predictions for TeV photon emission
derived from these models the HEGRA telescopes have observed Cas A for 234
hours in the observation periods from 1997 until 1999 under good conditions
of weather and detector. A crucial task in the analysis of data that have
been gathered over a prolonged period of time is the careful treatment of
the variations in detector performance.  \end{figwindow} The result of this
exceptionally deep observation and the careful analysis of the data is a
signal with a significance of \mbox{4.9 $\sigma$}.\cite{puehl} Different
analysis schemes applied to the same data have confirmed the signal.
Systematic checks (source position, energy spectrum of the excess events)
have shown that the observed excess is consistent with  a signal due to
photons (Fig. 1).

 The differential energy spectrum derived from the photon signal can be
characterized by a power law of the form $dN/dE \propto E^{-\alpha}$ with
$\alpha=2.5\pm0.1_{syst}\pm0.4_{stat}$ between energies of 1 and 10 TeV. The
spectral index slightly favors a nucleonic origin of the observed signal but
a leptonic origin can not be excluded.  
\section{Flaring activity of Mkn 421
during Feb-May 2000} \begin{figwindow}[1,l,{
\includegraphics[width=7cm]{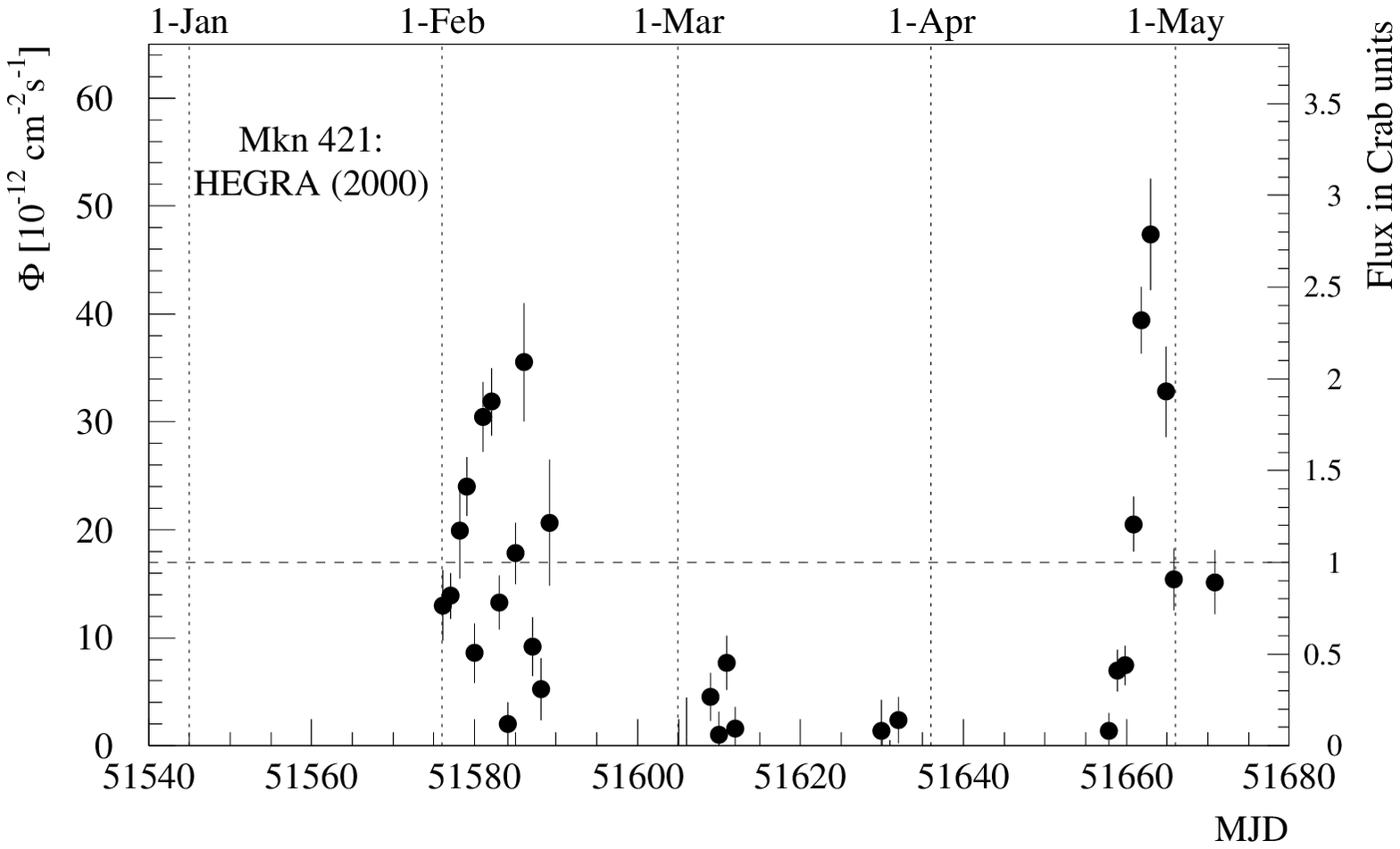}}, {As a function of modified
Julian date, the integral flux as measured by the HEGRA system of IACTs is
given. The averages are calculated for individual nights.
\label{figure:tev_lightcurve}}] The extragalactic sources Markarian 421 (Mkn
421) and Markarian 501 (Mkn 501) have been extensively studied in different
energy intervals.  These active galactic nuclei (AGN) of the so called
\textit{Blazar} type have shown strong variability over all observed energy
bands. Simultaneous observations in the X-ray and TeV energy bands reveal a
clear correlation of the time dependent fluxes. The observed spectral energy
distributions  and the correlated variability in X-ray and TeV emission of
Blazars can be explained in the framework of a synchrotron-self-Compton
model (SSC), where an energetic $e^+e^-$ plasma  moving with relativistic
bulk speed along the jet axis ($\Gamma=(1-\beta^2)^{-1/2}\approx10$ to $20$)
emits synchrotron photons with a peak in the spectral energy distribution at
energies of $0.1$ to $100$ keV in the observer's rest frame. By Compton
scattering of the energetic electrons off the synchrotron photons TeV
photons are produced.  As a consequence, the X-ray and TeV fluxes are
correlated.  \end{figwindow}

\begin{figwindow}[4,r,{
\includegraphics[width=7cm]{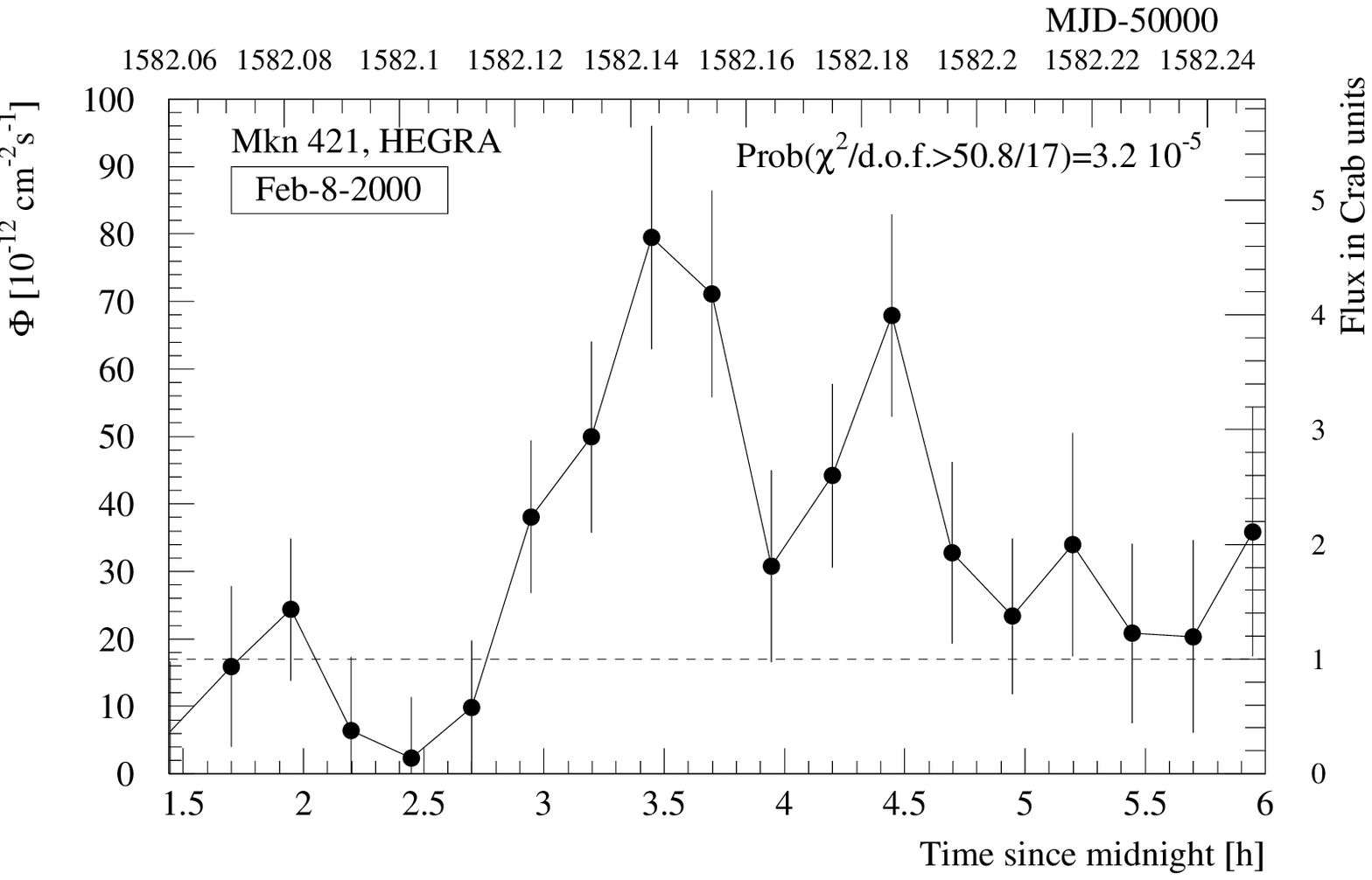}},
{ For MJD=51582 (Feb 8,2000), the integral fluxes for intervals of 15 minutes length  
   show a strong increase and a subsequent decrease in flux.\label{figure:tev_diurnal}}]
 The light curve of measured integral fluxes scaled 
to the flux measured from the Crab Nebula  above 1 TeV \cite{Konop_crab}
($\Phi_{Crab}(E>1\,\mbox{TeV})=1.7\cdot10^{-11}$ cm$^{-2}$s$^{-1}$) is displayed
in Fig. \ref{figure:tev_lightcurve} for the observational periods from February until May 2000.
% The results
%of a simultaneous observation in the X-ray energy band together with the RXTE and the BeppoSAX satellite will be
%described in forthcoming papers. 
The remarkable variability in February with strong indication
for intra-night variability is complementary to the rather slowly varying
TeV flux in the April and May data set.
As an example for the intra-night variability, Fig. \ref{figure:tev_diurnal} displays the 
integral fluxes obtained within 15 minutes time intervals on the 
8th of February 2000. The hypothesis of
a constant flux has a probability according to a $\chi^2$-test of $3.2\cdot 10^{-5}$.
 The TeV energy spectrum of Mkn 421 extends beyond 6 TeV and is well described by a power law. The diurnal energy 
spectra of April and May show indications for spectral hardening with increasing flux.  The energy spectrum derived
from the observation of the night with the highest detected flux
(April 28$^\mathrm{th}$/29$^\mathrm{th}$ 2000) can be described by a power law
with a photon index of $2.3\pm0.1$. The averaged energy spectrum of the observations in 1997 and 1998 with the
HEGRA instruments was softer with a photon index of $3.04\pm0.07$. \cite{mkn421_old}
\end{figwindow}

 The observations were coordinated with satellite borne experiments like
BeppoSAX and RXTE to
take data simultaneously in different energy regions.  The results of these campaigns will be
described in forthcoming publications.

\section{Conclusions}
 The stereoscopic technique has proven to be highly successful in terms of sensitivity and accuracy of shower
reconstruction. The unmatched flux sensitivity allowed the detection of Cas A, a weak source of 
TeV $\gamma$-rays after a deep exposure of 234 hours at the level of 3\,\% of the
flux from the Crab Nebula.
The light-curve of the extragalactic source Mkn 421 has been measured during two strong flaring periods in February and 
April/May 2000. The source showed a rapidly changing flux 
with strong indications for intra-night variations during the February
observation period. Within one night (April 28/29, 2000) 
Mkn 421 showed the largest flux so-far detected 
by the HEGRA instruments from this source with 
an exceptionally hard diurnal energy spectrum that can be well described by a power law with a photon index of 
$2.3\pm0.1$. 
% The peak flux in the night from 28 to 29th of April reached an average value of (3.1$\pm$0.2 ) Crab flux units.

\end{document}